\documentclass[12pt,psfig]{article}
\usepackage{graphicx}
\begin{document}

\begin{center} 
{\bf  Current status of                                                                 
 Dirac-Brueckner-Hartree-Fock calculations in asymmetric nuclear matter 
 }              
\end{center}
\vspace{0.1cm} 
\begin{center} 
 Francesca Sammarruca and Pei Liu \\ 
\vspace{0.2cm} 
 Physics Department, University of Idaho, Moscow, ID 83844, U.S.A   
\end{center} 
\begin{abstract}
We review the current status of our microscopic calculations in asymmetric nuclear matter.
Updated predictions of the equation of state are made available to potential users. 
We discuss the features of our EoS in relation to the predicted neutron star maximum masses. 
\\ \\ 
PACS number(s): 21.65.+f,21.30.Fe 
\end{abstract}

\section{Introduction} 
                                                                     
Microscopic predictions of the nuclear equation of state (EoS), together with empirical constraints 
from EoS-sensitive observables, are a powerful combination to learn about the in-medium
behavior of the nuclear force. With this objective in mind, 
over the past several years our group has taken a broad look at the EoS exploring diverse 
aspects and extreme states of nuclear matter.

The asymmetric matter predictions originally calculated in Ref.~\cite{AS03} and used as input 
in a neutron star calculation in 2006 \cite{KS06} are being refined in this paper. The revisions 
concern mainly some technical aspects of the (most problematic) high-density part of the calculation.   
We have improved the determination of the single-nucleon potential                                 
and the convergence of the high-density calculation. 
We provide extensive numerical tables for the EoS for both neutron and    
$\beta$-equilibrated matter, 
as well as updated neutron star maximum masses and radii predictions. These revised EoSs               
replace the previous ones published by our group.    

We take the opportunity to discuss 
parametrizations of the symmetry energy for the convenience of potential applications 
in nuclear reactions, which we encourage experimentalists to consider. Finally, we also   
review our most recent efforts and work in 
progress. 

\section{EoS and symmetry energy} 

In Tables~1 and 2, we provide the equation of state for neutron and $\beta$-equilibrated matter (protons,         
electrons, and muons), respectively, 
in the form of 
energy density and pressure as a function of the baryon density. (In the case of baryon-lepton matter, 
electrons are treated as extremely 
relativistic non-interacting fermions, whereas muons are handled non-relativistically.) 
The units in the Tables are chosen such that the values can be directly applied in the public software 
available at the website {\it http://www.gravity.phys.uwm.edu/rns},                
which we have used to calculate             
 neutron star properties. The Bonn B potential \cite{Mac89} is employed throughout. 

In Fig.~1, we show the revised EoS for symmetric matter (solid red) and neutron matter (dashed black). 
Our EoS's can be characterized as being moderately soft at low to medium density                      
(the saturation density and energy being equal to 0.185 fm$^{-3}$ and -16.1 MeV, respectively),
and  fairly ``stiff" at 
high densities This feature originates from the strongly density-dependent repulsion inherent to the 
Dirac-Brueckner-Hartee-Fock (DBHF) method.  
In Ref.~\cite{Fuchs}, it is pointed out 
that constraints from neutron star phenomenology together with flow data from heavy-ion (HI) 
reactions suggest that such EoS behavior may be desirable. 
We will come back to this point later, in conjunction with neutron star predictions. 

\begin{figure}
\begin{center}
\vspace*{-4.0cm}
\hspace*{-2.0cm}
\scalebox{0.4}{\includegraphics{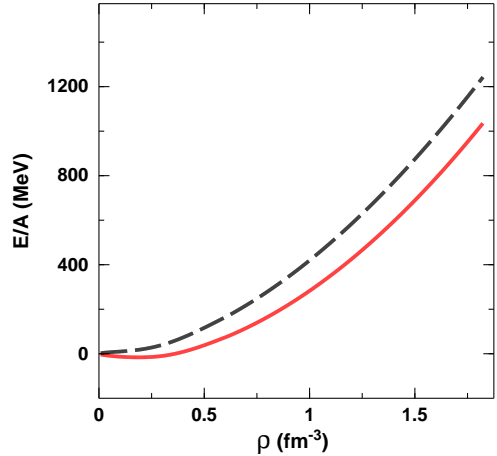}}
\vspace*{-1.5cm}
\caption{Updated predictions for the EoS of symmetric matter (solid red) and neutron matter (dashed black).            
} 
\label{one}
\end{center}
\end{figure}

\begin{table}
\scriptsize    
\vspace*{-3.0cm} 
\caption{Equation of state of pure neutron matter.} 
\begin{center}
\begin{tabular}{|c|c|c|}
\hline
Baryon density(1/$cm^3$)  & Energy density($g/cm^3$)  & Pressure($dyne/cm^2$)  \\                                         
                  \cline{1-3}
 0.675475E+23  &  0.113137E+00  &  0.458805E+07 \\
 0.540380E+24  &  0.905097E+00  &  0.175757E+09 \\
 0.432304E+25  &  0.724077E+01  &  0.697463E+10 \\
 0.145902E+26  &  0.244376E+02  &  0.612048E+11 \\
 0.345843E+26  &  0.579262E+02  &  0.295413E+12 \\
 0.675475E+26  &  0.113137E+03  &  0.873736E+12 \\
 0.540380E+27  &  0.905097E+03  &  0.304507E+14 \\
 0.432304E+28  &  0.724077E+04  &  0.124028E+16 \\
 0.145902E+29  &  0.244376E+05  &  0.108839E+17 \\
 0.345843E+29  &  0.579262E+05  &  0.525327E+17 \\
 0.675475E+29  &  0.113137E+06  &  0.155375E+18 \\
 0.540380E+30  &  0.905097E+06  &  0.541499E+19 \\
 0.432304E+31  &  0.724077E+07  &  0.220557E+21 \\
 0.145902E+32  &  0.244376E+08  &  0.193547E+22 \\
 0.345843E+32  &  0.579262E+08  &  0.960983E+22 \\
 0.675475E+32  &  0.113137E+09  &  0.209024E+23 \\
 0.844343E+34  &  0.141423E+11  &  0.107444E+27 \\
 0.675475E+35  &  0.113145E+12  &  0.476298E+28 \\
 0.227973E+36  &  0.381904E+12  &  0.422640E+29 \\
 0.540380E+36  &  0.905398E+12  &  0.196115E+30 \\
 0.105543E+37  &  0.176874E+13  &  0.640400E+30 \\
 0.182378E+37  &  0.305723E+13  &  0.167888E+31 \\
 0.289610E+37  &  0.485643E+13  &  0.378652E+31 \\
 0.432304E+37  &  0.725223E+13  &  0.765334E+31 \\
 0.615526E+37  &  0.103309E+14  &  0.146496E+32 \\
 0.844343E+37  &  0.141791E+14  &  0.226342E+32 \\
 0.145902E+38  &  0.245257E+14  &  0.434992E+32 \\
 0.231688E+38  &  0.389881E+14  &  0.874118E+32 \\
 0.345843E+38  &  0.582634E+14  &  0.156362E+33 \\
 0.492421E+38  &  0.830558E+14  &  0.257664E+33 \\
 0.675475E+38  &  0.114075E+15  &  0.468635E+33 \\
 0.781946E+38  &  0.132151E+15  &  0.621198E+33 \\
 0.899057E+38  &  0.152061E+15  &  0.816935E+33 \\
 0.102731E+39  &  0.173905E+15  &  0.111487E+34 \\
 0.116722E+39  &  0.197791E+15  &  0.156721E+34 \\
 0.131929E+39  &  0.223833E+15  &  0.223210E+34 \\
 0.148402E+39  &  0.252158E+15  &  0.319807E+34 \\
 0.166192E+39  &  0.282905E+15  &  0.463315E+34 \\
 0.185350E+39  &  0.316247E+15  &  0.681545E+34 \\
 0.205927E+39  &  0.352381E+15  &  0.993009E+34 \\
 0.227973E+39  &  0.391534E+15  &  0.141505E+35 \\
 0.251538E+39  &  0.433969E+15  &  0.203050E+35 \\
 0.276674E+39  &  0.480078E+15  &  0.293314E+35 \\
 0.303432E+39  &  0.530247E+15  &  0.409031E+35 \\
 0.331861E+39  &  0.584918E+15  &  0.558832E+35 \\
 0.362012E+39  &  0.644589E+15  &  0.746722E+35 \\
 0.393937E+39  &  0.709762E+15  &  0.965629E+35 \\
 0.427685E+39  &  0.780907E+15  &  0.122694E+36 \\
 0.463308E+39  &  0.858621E+15  &  0.152765E+36 \\
 0.500856E+39  &  0.943372E+15  &  0.185795E+36 \\
 0.540380E+39  &  0.103570E+16  &  0.223317E+36 \\
 0.581930E+39  &  0.113630E+16  &  0.268106E+36 \\
 0.625557E+39  &  0.124615E+16  &  0.324163E+36 \\
 0.671312E+39  &  0.136650E+16  &  0.395072E+36 \\
 0.719245E+39  &  0.149890E+16  &  0.485417E+36 \\
 0.769408E+39  &  0.164502E+16  &  0.589746E+36 \\
 0.821850E+39  &  0.180610E+16  &  0.707244E+36 \\
 0.876622E+39  &  0.198388E+16  &  0.846053E+36 \\
 0.933776E+39  &  0.218030E+16  &  0.100924E+37 \\
 0.993361E+39  &  0.239765E+16  &  0.120127E+37 \\
 0.105543E+40  &  0.263842E+16  &  0.142307E+37 \\
 0.112003E+40  &  0.290521E+16  &  0.167838E+37 \\
 0.118721E+40  &  0.320102E+16  &  0.197388E+37 \\
 0.125703E+40  &  0.352921E+16  &  0.231442E+37 \\
 0.132954E+40  &  0.389345E+16  &  0.270588E+37 \\
 0.140478E+40  &  0.429785E+16  &  0.315475E+37 \\
 0.148280E+40  &  0.474691E+16  &  0.366826E+37 \\
 0.156366E+40  &  0.524560E+16  &  0.425495E+37 \\
 0.164741E+40  &  0.579940E+16  &  0.492059E+37 \\
 0.173410E+40  &  0.641418E+16  &  0.567978E+37 \\
 0.182378E+40  &  0.709679E+16  &  0.654219E+37 \\
\hline
\end{tabular}
\end{center}
\end{table}

\begin{table}
\scriptsize   
\vspace*{-3cm}
\caption{As in the previous Table, for $\beta$-equilibrated matter.}                                       
\begin{center}
\begin{tabular}{|c|c|c|}
\hline
Baryon density(1/$cm^3$)  & Energy density($g/cm^3$)  & Pressure($dyne/cm^2$)  \\                                         
                  \cline{1-3}
 0.675475E+23  &  0.113137E+00  &  0.458805E+07 \\
 0.540380E+24  &  0.905097E+00  &  0.175757E+09 \\
 0.432304E+25  &  0.724077E+01  &  0.697463E+10 \\
 0.145902E+26  &  0.244376E+02  &  0.612048E+11 \\
 0.345843E+26  &  0.579262E+02  &  0.295413E+12 \\
 0.675475E+26  &  0.113137E+03  &  0.873736E+12 \\
 0.540380E+27  &  0.905097E+03  &  0.304507E+14 \\
 0.432304E+28  &  0.724077E+04  &  0.124028E+16 \\
 0.145902E+29  &  0.244376E+05  &  0.108839E+17 \\
 0.345843E+29  &  0.579262E+05  &  0.525327E+17 \\
 0.675475E+29  &  0.113137E+06  &  0.155375E+18 \\
 0.540380E+30  &  0.905097E+06  &  0.541499E+19 \\
 0.432304E+31  &  0.724077E+07  &  0.220557E+21 \\
 0.145902E+32  &  0.244376E+08  &  0.193547E+22 \\
 0.345843E+32  &  0.579262E+08  &  0.960983E+22 \\
 0.675475E+32  &  0.113137E+09  &  0.209024E+23 \\
 0.844343E+34  &  0.141423E+11  &  0.107418E+27 \\
 0.675475E+35  &  0.113145E+12  &  0.477163E+28 \\
 0.227973E+36  &  0.381904E+12  &  0.422897E+29 \\
 0.540380E+36  &  0.905397E+12  &  0.196017E+30 \\
 0.105543E+37  &  0.176874E+13  &  0.639887E+30 \\
 0.182378E+37  &  0.305723E+13  &  0.167466E+31 \\
 0.289610E+37  &  0.485641E+13  &  0.376394E+31 \\
 0.432304E+37  &  0.725216E+13  &  0.756110E+31 \\
 0.615526E+37  &  0.103307E+14  &  0.143581E+32 \\
 0.844343E+37  &  0.141786E+14  &  0.216127E+32 \\
 0.145902E+38  &  0.245236E+14  &  0.416322E+32 \\
 0.231688E+38  &  0.389827E+14  &  0.822804E+32 \\
 0.345843E+38  &  0.582501E+14  &  0.142104E+33 \\
 0.492421E+38  &  0.830269E+14  &  0.230061E+33 \\
 0.675475E+38  &  0.114019E+15  &  0.409814E+33 \\
 0.781946E+38  &  0.132073E+15  &  0.540126E+33 \\
 0.899057E+38  &  0.151955E+15  &  0.711724E+33 \\
 0.102731E+39  &  0.173765E+15  &  0.970498E+33 \\
 0.116722E+39  &  0.197605E+15  &  0.137046E+34 \\
 0.131929E+39  &  0.223589E+15  &  0.201627E+34 \\
 0.148402E+39  &  0.251837E+15  &  0.261928E+34 \\
 0.166192E+39  &  0.282481E+15  &  0.330297E+34 \\
 0.185350E+39  &  0.315683E+15  &  0.517205E+34 \\
 0.205927E+39  &  0.351619E+15  &  0.757028E+34 \\
 0.227973E+39  &  0.390495E+15  &  0.110037E+35 \\
 0.251538E+39  &  0.432555E+15  &  0.160843E+35 \\
 0.276674E+39  &  0.478138E+15  &  0.235092E+35 \\
 0.303432E+39  &  0.527601E+15  &  0.334557E+35 \\
 0.331861E+39  &  0.581355E+15  &  0.465834E+35 \\
 0.362012E+39  &  0.639861E+15  &  0.632143E+35 \\
 0.393937E+39  &  0.703594E+15  &  0.832228E+35 \\
 0.427685E+39  &  0.773030E+15  &  0.107139E+36 \\
 0.463308E+39  &  0.848703E+15  &  0.134775E+36 \\
 0.500856E+39  &  0.931088E+15  &  0.165606E+36 \\
 0.540380E+39  &  0.102069E+16  &  0.200271E+36 \\
 0.581930E+39  &  0.111814E+16  &  0.241417E+36 \\
 0.625557E+39  &  0.122434E+16  &  0.293125E+36 \\
 0.671312E+39  &  0.134047E+16  &  0.358691E+36 \\
 0.719245E+39  &  0.146797E+16  &  0.442173E+36 \\
 0.769408E+39  &  0.160841E+16  &  0.539098E+36 \\
 0.821850E+39  &  0.176298E+16  &  0.649283E+36 \\
 0.876622E+39  &  0.193337E+16  &  0.779667E+36 \\
 0.933776E+39  &  0.212141E+16  &  0.932987E+36 \\
 0.993361E+39  &  0.232930E+16  &  0.111395E+37 \\
 0.105543E+40  &  0.255941E+16  &  0.132452E+37 \\
 0.112003E+40  &  0.281437E+16  &  0.156916E+37 \\
 0.118721E+40  &  0.309721E+16  &  0.185697E+37 \\
 0.125703E+40  &  0.341144E+16  &  0.219046E+37 \\
 0.132954E+40  &  0.376060E+16  &  0.257227E+37 \\
 0.140478E+40  &  0.414881E+16  &  0.301377E+37 \\
 0.148280E+40  &  0.458067E+16  &  0.352202E+37 \\
 0.156366E+40  &  0.506130E+16  &  0.410655E+37 \\
 0.164741E+40  &  0.559636E+16  &  0.477543E+37 \\
 0.173410E+40  &  0.619202E+16  &  0.554149E+37 \\
 0.182378E+40  &  0.685522E+16  &  0.641233E+37 \\
\hline
\end{tabular}
\end{center}
\end{table}

In Fig.~2, we display our DBHF predictions for the symmetry energy, solid red curve. 
The latter is seen to grow at a lesser rate with increasing density, 
 an indication that, at large density,   
repulsion in the symmetric matter EoS increases at 
rapid rate relative to the neutron matter EoS.   
This can be understood in terms of increased repulsion in isospin zero partial waves (absent
from neutron matter) as a function of density, see Table.~3. 
Our predicted value for the symmetry pressure, $L$, is 69.6 MeV. 

The various black dashed curves in Fig.~2 are obtained with the simple parametrization               
\begin{equation}
e_{sym}=C(\rho/\rho_0)^{\gamma}, 
\end{equation}
with $\gamma$=0.7-1.0, and $C \approx 32$ MeV. 
It seems that a value of $\gamma$ close to 0.8 gives a reasonable description of our predictions, 
although the use of different functions in different density regions may be best for an 
optimal fit. This can easily be done upon request, with an eye on the particular density region to be
probed by the experiment.  

Considering that all of the dashed curves are 
commonly used parametrizations 
suggested by HI data \cite{BA}, 
Fig.~2 clearly reflects the large uncertainty in our knowledge of the symmetry energy at 
the larger densities (in fact, already above 2-3$\rho_0$).

\begin{figure}
\begin{center}
\vspace*{-4.0cm}
\hspace*{-2.0cm}
\scalebox{0.4}{\includegraphics{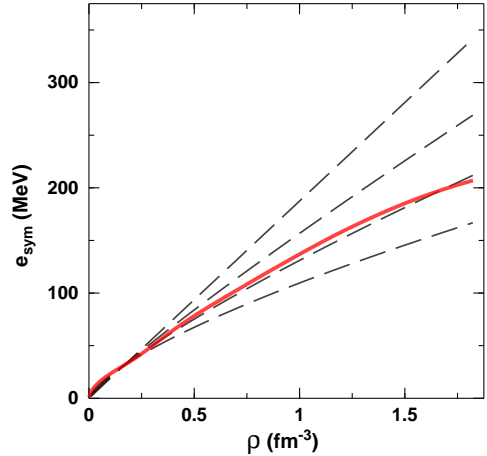}}
\vspace*{-3.0cm}
\caption{DBHF predictions for the symmetry energy (solid red) compared with various         
phenomenological parametrizations (dashed black). See text for details. 
} 
\label{two}
\end{center}
\end{figure}
\begin{table}
\caption{Contribution to the potential energy of nuclear matter (in MeV) from some isospin zero
partial waves. 
} 
\begin{center}
\begin{tabular}{|c|c|c|}
\hline
$\rho(fm^{-3})$   &   $^3S_1$  &    $^1P_1$ \\             
                  \cline{1-3}
 0.1853  &         -20.30              & 4.044 \\                                  
 0.5404  &         -14.87              & 15.72 \\                                  
 1.0554  &          8.676              & 36.33 \\                                  
\hline
\end{tabular}
\end{center}
\end{table}

\begin{table}
\caption{ Neutron star maximum masses, radii, and corresponding central densities. 
} 
\begin{center}
\begin{tabular}{|c|c|c|c|}
\hline
Matter type & $M_{max}(M_{\odot})$  & $R_{M_{max}}$(km) & Central density(g/cm$^{3}$) \\             
                  \cline{1-4}
neutrons &           2.33              & 11.2 & 2.17$\times$ 10$^{15}$         \\ 
                  \cline{1-4}
$\beta$-equilibrated    &  2.27 & 10.8       & 2.26$\times$ 10$^{15}$                       \\ 
\hline
\end{tabular}
\end{center}
\end{table}

\section{EoS and neutron star predictions}                                
In Table 3, we report our latest predictions of the basic neutron star properties, which 
are well within the (rather large) range covered by most realistic models. 
We will try to get some deeper insight from a closer look at 
some of the predictions included in the analysis of Ref.~\cite{Fuchs}, such as 
relativistic mean field (RMF) models. Examples are those from 
 Refs.~\cite{DD,D3C}, which use density-dependent (``DD") meson couplings and are 
fitted to the properties of nuclei up to about 0.15 fm$^{-3}$. They generate the steepest 
EoSs and thus the largest pressure.                                                                      
An improvement to the traditional RMF description of nuclear matter 
can be obtained through the introduction of non-linear (``NL") self-interactions of the $\sigma$ meson, such as done 
in the models of Refs.~\cite{NL1,NL2}, with the parametrization of Ref.~\cite{NL2} including 
the $\delta$ meson in addition to the usual $\rho$. The corresponding EoSs are much less repulsive
than those of ``DD" models (although the symmetry energy becomes very large at high 
density, possibly due to the absence of non-linearity and density dependence at the 
isovector level). 

Clearly, the pressure as a function of density plays the crucial role in building the structure of the
star. 
In Fig.~3 we show our predicted pressure in symmetric matter compared with constraints obtained
from flow data \cite{MSU}. 
The predictions are seen to fall just on the high side of the constraints and grow
rather steep at high density.                     
Comparing with 
Fig.~6 of Ref.~\cite{Fuchs}, we see that                                                      
our predictions                                                                      
are well below those of DD-RMF models at low to 
moderate density but nearly catch up with them at very high density, a description that    
would also be appropriate for the 
DBHF predictions of Ref.~\cite{Fuchs2} (red curve in Fig.~6 of Ref.~\cite{Fuchs}). 
Of all the cases studied in 
Ref.~\cite{Fuchs}, DD-RMF models predict the largest maximum masses and radii and the 
lowest central densities. Thus,           
an equation of state where high pressure is sustained for a longer radial distance (moving away 
from the center of the star) 
will allow the maximum mass star to be heavier, larger, and more ``diffuse" at the center. 
On the other hand, 
microscopic relativistic models, (such as the DBHF calculation of Ref.~\cite{Fuchs2}
or the present one, which are in reasonable agreement with each other) display a rather different              
density dependence of the pressure and produce smaller and more compact maximum mass stars. 
(All other EoS's considered in Ref.~\cite{Fuchs} are softer and generate smaller maximum masses 
with smaller radii and larger central densities.)                           

To conclude this Section, we show in Fig.~4 the predicted pressure in neutron matter (red curve)
and $\beta$-equilibrated matter (green). The pressure contour is again from Ref.~\cite{MSU} 
and was obtained from flow data together with the assumption of strong density dependence in 
the asymmetry term (indicated as ``Asy\_ stiff'' in Ref.~\cite{MSU}). 

\begin{figure}
\begin{center}
\vspace*{-4.0cm}
\hspace*{-2.0cm}
\scalebox{0.4}{\includegraphics{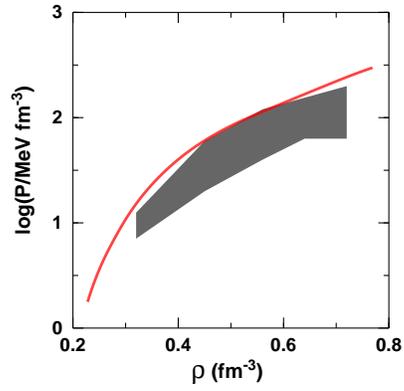}}
\vspace*{-1.5cm}
\caption{Pressure in symmetric matter. The shaded area corresponds to the region
of pressure consistent with the flow data analysed in Ref.~\cite{MSU}.} 
\label{three}
\end{center}
\end{figure}

\begin{figure}
\begin{center}
\vspace*{-4.0cm}
\hspace*{-2.0cm}
\scalebox{0.4}{\includegraphics{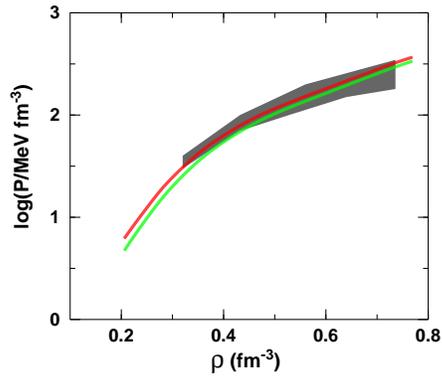}}
\vspace*{-1.5cm}
\caption{Pressure in neutron (red curve) and baryon-lepton (green curve) matter. The             
shaded area corresponds to the region
of pressure consistent with flow data and the inclusion of strong                            
density dependence in the asymmetry terms \cite{MSU}.} 
\label{four}
\end{center}
\end{figure}

\section{Outlook}                                                                  
The revised EoS presented in this paper is  the 
baseline for inclusion of species other than nucleons and leptons, e.g., hyperons; cf. Ref.~\cite{Sam08}.

At the same time, we continue to explore, broadly, several EoS-related aspects. 
Microscopic calculations of the EoS generate, in an internally consistent manner, 
quantities such as the single-nucleon potential and thus effective masses. In turn, those
enter the calculation of in-medium cross sections and the nucleon mean free path \cite{MFP}. 
Our goal                                                                                
is to consistently predict and critically analyse all of these EoS-dependent                                 
``observables", 
looking for patterns and characteristic signatures which, together with available
constraints, may help us identify model strengths and weaknesses. 
Ultimately, 
coherent effort from theory, experiment, phenomenology, and observations will be 
essential to improve our knowledge of such fundamentally important quantity as the 
nuclear equation of state.

\section*{Acknowledgments}
Support from the U.S. Department of Energy under Grant No. DE-FG02-03ER41270 is 
acknowledged.                                                                           
%\section{References}

\end{document}